\documentclass[twocolumn,showpacs,preprintnumbers,superscriptaddress,amsmath,amssymb]{revtex4}

\usepackage{amssymb}
\usepackage{amsmath}
\usepackage{graphicx}
\usepackage[normalem]{ulem}
\usepackage{multirow}
\usepackage{mhchem}
\usepackage{CJK}
\usepackage{appendix}
\usepackage[colorlinks, citecolor=blue, linkcolor=blue, linkcolor=blue]{hyperref}
\usepackage[figuresright]{rotating}

\begin{document}
\begin{CJK*} {UTF8} {gbsn}

\title{Neutron capture measurement and resonance parameter analysis of \ce{^{nat}Sm}}

\author{X.X. Li (李鑫祥)}
\affiliation{Shanghai Institute of Applied Physics, Chinese Academy
of Sciences, Shanghai 201800, China}
\affiliation{University of Chinese Academy of Sciences, Beijing 100049, China}
\author{ L.X. Liu (刘龙祥)}
\affiliation{Shanghai Advanced Research Institute, Chinese Academy of Sciences,
Shanghai 201210, China}
\affiliation{Shanghai Institute of Applied Physics, Chinese Academy
of Sciences, Shanghai 201800, China}
\author{ W. Jiang (蒋伟)}
\affiliation{Institute of High Energy Physics, Chinese Academy of Sciences, Beijing,100049,China}
\affiliation{China Spallation Neutron Source, Dongguan, 523803, China}
\author{ J. Ren (任杰)}
\affiliation{China Institute of Atomic Energy, Beijing, 102413, China}
\author{H.W. Wang (王宏伟)\footnote{Corresponding author: wanghongwei@zjlab.org.cn}}
\affiliation{Shanghai Advanced Research Institute, Chinese Academy of Sciences,
Shanghai 201210, China}
\affiliation{Shanghai Institute of Applied Physics, Chinese Academy
of Sciences, Shanghai 201800, China}
\affiliation{University of Chinese Academy of Sciences, Beijing 100049, China}
\author{G.T. Fan (范功涛)\footnote{Corresponding author: fangongtao@zjlab.org.cn}}
\affiliation{Shanghai Advanced Research Institute, Chinese Academy of Sciences,
Shanghai 201210, China}
\affiliation{Shanghai Institute of Applied Physics, Chinese Academy
of Sciences, Shanghai 201800, China}
\affiliation{University of Chinese Academy of Sciences, Beijing 100049, China}
\author{J.J. He (何建军)}
\affiliation{Key Laboratory of Beam Technology of Ministry of Education, College of Nuclear Science and Technology, Beijing Normal University, Beijing 100875, China}
\affiliation{Institute of Radiation Technology, Beijing Academy of Science and Technology, Beijing 100875, China}
\author{ D.X. Wang (王德鑫)}
\affiliation{College of Mathematics and Physics, Inner Mongolia Minzu University , Tongliao 028000,China}
\author{S.Y Zhang (张苏雅拉吐)}
\affiliation{College of Mathematics and Physics, Inner Mongolia Minzu University , Tongliao 028000,China}
\author{ Z.D. An (安振东)}
\affiliation{Shanghai Institute of Applied Physics, Chinese Academy
of Sciences, Shanghai 201800, China}
\affiliation{Sun Yat-sen University, Zhuhai, 510275, China}
\author{ X.G. Cao (曹喜光)}
\affiliation{Shanghai Advanced Research Institute, Chinese Academy of Sciences,
Shanghai 201210, China}
\affiliation{Shanghai Institute of Applied Physics, Chinese Academy
of Sciences, Shanghai 201800, China}
\affiliation{University of Chinese Academy of Sciences, Beijing 100049, China}
\author{ L.L. Song (宋龙龙)}
\affiliation{Shanghai Advanced Research Institute, Chinese Academy of Sciences,
Shanghai 201210, China}
\author{Y. Zhang (张岳)}
\affiliation{Institute of High Energy Physics, Chinese Academy of Sciences, Beijing,100049,China}
\affiliation{China Spallation Neutron Source, Dongguan, 523803, China}
\author{X.R. Hu (胡新荣)}
\affiliation{Shanghai Institute of Applied Physics, Chinese Academy
of Sciences, Shanghai 201800, China}
\affiliation{University of Chinese Academy of Sciences, Beijing 100049, China}
\author{ Z.R. Hao (郝子锐)}
\affiliation{Shanghai Institute of Applied Physics, Chinese Academy
of Sciences, Shanghai 201800, China}
\affiliation{University of Chinese Academy of Sciences, Beijing 100049, China}
\author{ P. Kuang (匡攀)}
\affiliation{Shanghai Institute of Applied Physics, Chinese Academy
of Sciences, Shanghai 201800, China}
\affiliation{University of Chinese Academy of Sciences, Beijing 100049, China}
\author{ B. Jiang (姜炳)}
\affiliation{Shanghai Institute of Applied Physics, Chinese Academy
of Sciences, Shanghai 201800, China}
\affiliation{University of Chinese Academy of Sciences, Beijing 100049, China}
\author{ X.H. Wang (王小鹤)}
\affiliation{Shanghai Institute of Applied Physics, Chinese Academy
of Sciences, Shanghai 201800, China}
\author {J.F. Hu (胡继峰)}
\affiliation{Shanghai Institute of Applied Physics, Chinese Academy
of Sciences, Shanghai 201800, China}
\author{Y.D. Liu (刘应都)}
\affiliation{Xiangtan University, Xiangtan, 411105, China}
\author{X. Ma (麻旭)}
\affiliation{Xiangtan University, Xiangtan, 411105, China}
\author{C.W. Ma (马春旺)}
\affiliation{Henan Normal University, Xinxiang, 453007, China}
\author{Y.T. Wang (王玉廷)}
\affiliation{Henan Normal University, Xinxiang, 453007, China}
\author{J. Su (苏俊)} 
\affiliation{Key Laboratory of Beam Technology and Material Modification of Ministry of Education, College of Nuclear Science and Technology, Beijing Normal University, Beijing 100875, China}
\affiliation{Beijing Radiation Center, Beijing 100875, China}
\author{ L.Y. Zhang (张立勇)}
\affiliation{Key Laboratory of Beam Technology and Material Modification of Ministry of Education, College of Nuclear Science and Technology, Beijing Normal University, Beijing 100875, China}
\affiliation{Beijing Radiation Center, Beijing 100875, China}
\author{ Y.X. Yang (杨宇萱)}
\affiliation{Shanghai Institute of Applied Physics, Chinese Academy
of Sciences, Shanghai 201800, China}
\author{ W.B. Liu (刘文博)}
\affiliation{Shanghai Institute of Applied Physics, Chinese Academy
of Sciences, Shanghai 201800, China}
\affiliation{Henan Normal University, Xinxiang, 453007, China}
\author{ W.Q. Su (苏琬晴)}
\affiliation{Shanghai Institute of Applied Physics, Chinese Academy
of Sciences, Shanghai 201800, China}
\affiliation{Henan Normal University, Xinxiang, 453007, China}
\author{ S. Jin (金晟)}
\affiliation{Shanghai Institute of Applied Physics, Chinese Academy
of Sciences, Shanghai 201800, China}
\affiliation{University of Chinese Academy of Sciences, Beijing 100049, China}
\author{ K.J. Chen (陈开杰)}
\affiliation{Shanghai Institute of Applied Physics, Chinese Academy
of Sciences, Shanghai 201800, China}
\affiliation{ShanghaiTech University, Shanghai 200120, China}

\date{\today}

\begin{abstract}

Multiple isotopes of samarium element are the isotopes produced by the $s$ process, and \ce{^{154}Sm} is produced by the $r$ process. In addition,  \ce{^{144}Sm} is $p$ nuclei in nuclear astrophysics. The measurement of these can help us to better understand the results of relevant photonuclear reaction experiments. On the other hand, \ce{^{149}Sm} is  a \ce{^{235}U} fission product with a $1\%$ yield,  its cross sections are important to reactor neutronics. In this work, the neutron capture yield of the natural samarium target was measured at the back-streaming white neutron beamline (Back-n) of the China Spallation Neutron Source (CSNS), and the resonance parameters were analyzed by \footnotesize{SAMMY} \normalsize code. The resonance peaks and the neutron separation energies contributed by the different isotopes are considered individually. The results of the capture yield found signs of the possibility of two resonance peaks at 8 eV, which awaits further experimental examination. Cross-section was calculated according to resonance parameters and was compared with other experimental results and evaluation databases of ENDF/B-VIII.0 and CENDL-3.2. 
A clear difference between ENDF/B VIII.0 and CENDL-3.2 database appears at 23.2 eV, the experimental result at this energy is smaller than data of ENDF/B VIII.0 database but CENDL-3.2 database. Most of the controversial experimental results invariably come from the samarium 149 isotope.

\end{abstract}

\maketitle

\section{INTRODUCTION}
\label{introduction}

The origin of heavy elements has always been a concern in physics. The elements from carbon to iron can be well explained by the nuclear combustion process of stars. However, these processes cannot produce nuclei with mass number $A>64$ \cite{bib.01}. Neutron capture reaction is the main way to synthesize elements heavier than iron, neutron capture processes in astrophysics include the slow neutron capture process ($s$ process) \cite{bib.a1} and the rapid neutron capture process ($r$ process) \cite{bib.a2}, \ce{^{147,148,149,150,152}Sm} are the isotopes produced by the $s$ process, and \ce{^{154}Sm} is produced by the $r$ process. Their neutron capture cross sections are of great significance for understanding the pathways taken during the $s$ process. In addition, about $1\%$ of heavy element abundances are produced by charged particles and photoinduced reactions ($p$ process) \cite{bib.a3}, \ce{^{144}Sm} is $p$ nuclei in nuclear astrophysics. It is not produced by neutron capture but rather carry out neutron capture reactions, the measurement of which can help us to better understand the results of relevant photonuclear reaction experiments. On the other hand, \ce{^{235}U} is an important raw material for nuclear reactors and samarium 149 isotope is  a \ce{^{235}U} fission product with a $1\%$ yield,  its cross sections are important to reactor neutronics.  

Few past studies on samarium neutron capture cross sections have focused on 1-50 eV \cite{bib.b1,bib.b2,bib.b3,bib.b5,bib.b6,bib.b7,bib.b8,bib.b9}. A neutron capture experiment on natural samarium was the work of Chou et al. in 1973 \cite{bib.cjc}. Unfortunately, the resonance parameters for one peak in the resolvable resonance energy region were not analyzed.   The latest results came from Leinweber et al \cite{bib.gl}. 20 years ago, they measured the neutron capture cross-section of samarium by using the dilute samples of samarium nitrate in deuterated water (\ce{D_2O}) at the Rensselaer Polytechnic Institute (RPI) LINAC facility, with neutron energies up to 30 eV. 

China spallation neutron source (CSNS) is the first spallation neutron source in China, and it has important applications in the fields of materials science, physics, chemistry, life sciences, resources and environment, and new energy \cite{CSNS1,CSNS2,CSNS3}. The back-streaming white neutron beamline (Back-n) was built in 2018, and it is mainly used for neutron data measurements \cite{backn.01, backn.02}. So far, a number of neutron capture reaction experiments have been carried out on the Back-n beamline \cite{lxx.hjs,hxr.nst,rj.cpc,wdx.aps,jb.cpb,lxx.prc,lxx.cpb}. 

This work reports the neutron capture cross-section of natural samarium in the energy region between 1 eV and 50 eV measured at the Back-n beamline. Resonance parameters are extracted by \footnotesize{SAMMY} \normalsize  code for each peak contributed by individual isotope in this energy region. The experimental methods, data processing and uncertainty analysis are presented.

\section{Experimental Method}
\label{experiment}

\begin{figure}[ht]
\centering
\includegraphics[width=.5\textwidth]{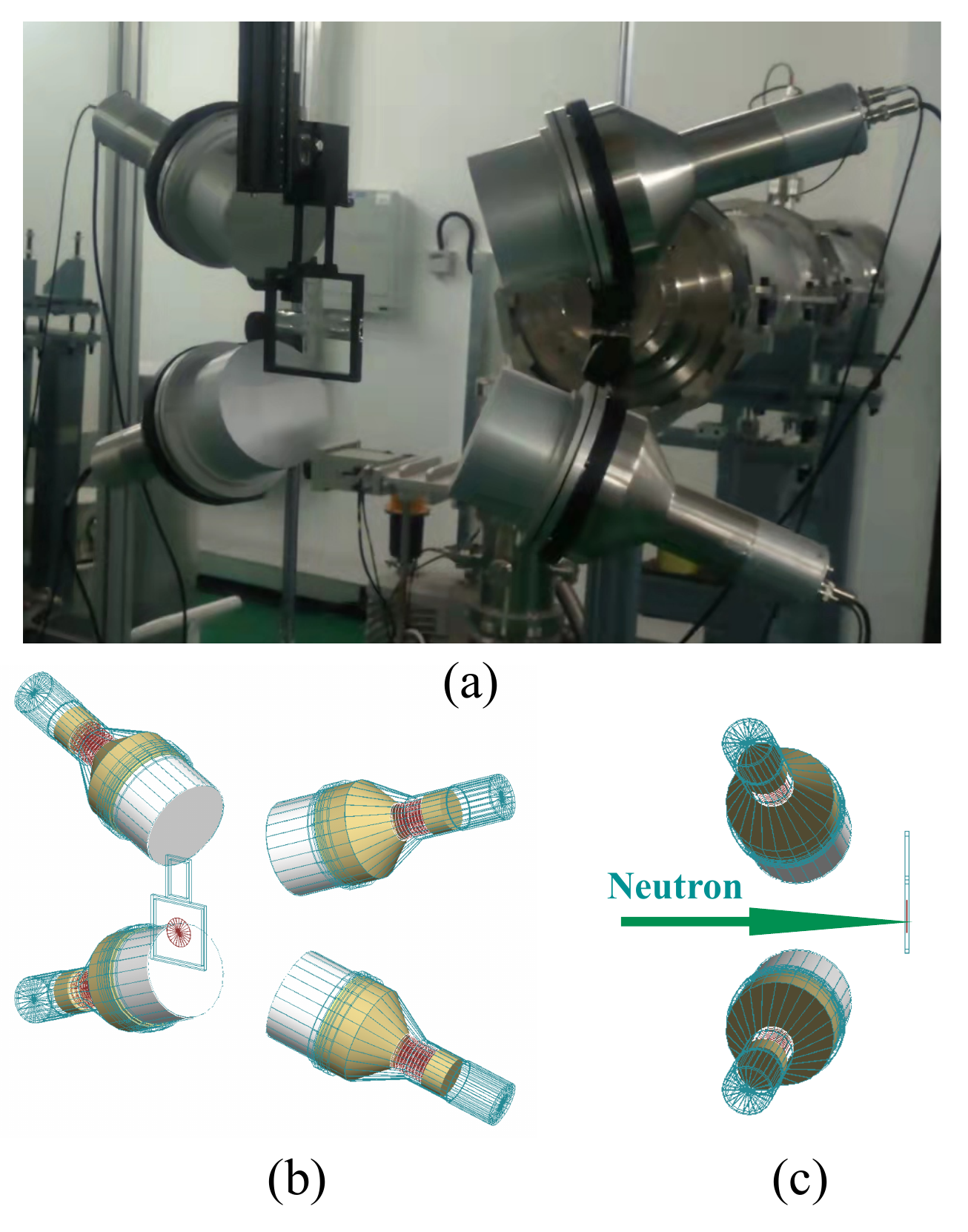}
\caption{(color online)   (a) A picture of the detector layout. (b) Monte Carlo simulation reconstruction of the detector system and target. (c) A side view of the layout.}
\label{Fig.021}
\end{figure}

The neutron capture experiment was carried out at the end station 2 (ES\#2) of the Back-n beamline. The measurement uses the \ce{C_6D_6} detection system. It consists of four \ce{C_6D_6} scintillation detectors, approximately 76 m away from the spallation target.  Each \ce{C_6D_6} scintillator has 127 mm in diameter and 76.2 mm in length, and contained in a 1.5-mm thick aluminum capsule which was coupled with a photomultiplier tube of ETEL 9390 KEB PMT \cite{rj.8}. The physical layout and Monte-Carlo reconstruction of the detector system and target are shown in Fig. \ref{Fig.021}(a) and (b) ( see Ref. \cite{lxx.prc} for details). Detector placement is backwards relative to the beam direction as shown in Fig. \ref{Fig.021}(c). This is because the $\gamma$ rays emitted from the ($n,\gamma$) reaction are isotropic, and the backward layout can effectively avoid the influence of neutron scattering without significantly reducing the detection efficiency. The neutron flux was measured by a Li-Si detector at neutron trash, which is based on the \ce{^{6}Li}($n, \alpha$)\ce{^{3}H} reaction. Here, the used neutron spectrum was provided by the Back-n collaboration \cite{cyh.epja}. Back-n data acquisition system (DAQ) adopts a full waveform data acquisition solution.

\begin{figure}[ht]
\centering
\includegraphics[width=.5\textwidth]{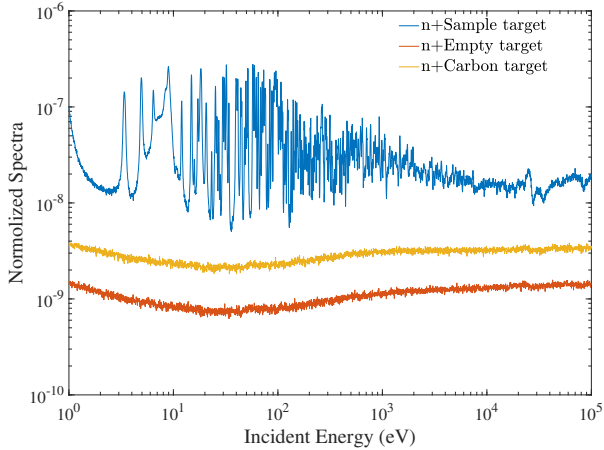}
\caption{(color online)  Raw spectra for different targets, normalized to the Li-Si detector results.}
\label{Fig.022}
\end{figure}

The experiment was carried out in May 2019. A gold (\ce{^{197}Au}) target, a carbon (\ce{^{nat}C}) target, an empty target and a natural samarium target are prepared. The total beam time was about 49 h. The \ce{^{197}Au} ($n,\gamma$) \ce{^{198}Au} reaction standard neutron capture was firstly measured for 13 h in proton power between 50.5 and 51.9 kW to reproduce the previous results \cite{lxx.hjs}. Carbon and empty targets were used to determine the background of neutron scattering and environmental background under beam conditions, and they were measured for 12 h and 8 h, respectively. During this period, the accelerator was relatively stable and the beam power was about 50 kW, with an uncertainty less than $2.0\%$. Finally, the natural samarium target was measured for 16 h in beam power of 48.28--50.45 kW. Normalized spectra for the 1 eV--100 keV energy region are shown in Fig. \ref{Fig.022}. The parameters of the targets and measurement conditions are shown in Table \ref{Tab.01}, where diameter was measured by vernier caliper and thickness was measured by micrometer. 

\begin{table*}[ht]\small
	\caption{Information of experimental targets}
	\begin{tabular}{c| c  c c| c |c | c }
	   \hline \hline
	   Target&  & Impurities  & & Diameter (mm) &Thickness  (mm)& Beam Power (kW)\\
	   \hline
	   \multirow{6}{*}{\ce{^{nat}Sm}}  
	   & $\omega(\ce{Mo}) = 0.002 \%$ & $ \omega (\ce{Ti}) = 0.002 \%$  & $\omega(\ce{Tb}) = 0.001 \%$ & \multirow{6}{*}{$50.00\pm0.02$} & \multirow{6}{*}{$1.000\pm 0.005$ } & \multirow{6}{*}{$49.37\pm 1.08$ }\\
	   & $ \omega(\ce{Fe}) = 0.01\%$ & $\omega(\ce{Ca}) = 0.005 \%$ & $\omega(\ce{C}) = 0.01 \%$ & & \\
	   & $\omega (\ce{si}) = 0.01 \%$ &  $\omega(\ce{Mg}) = 0.005 \%$ & $\omega(\ce{Nb})=0.002 \%$ & & \\
	   & $\omega(\ce{Al}) = 0.005 \%$ & $\omega(\ce{Cl}) = 0.005 \%$ & $\omega(\ce{Ta}) = 0.002 \%$ &  & \\
	   & $\omega(\ce{La}) = 0.001 \%$ & $\omega(\ce{Ce})= 0.001 \%$ & $\omega(\ce{Pr}) = 0.002 \%$ &  & \\
	        	        \hline
		        \ce{^{nat}C}  & & $ < 0.100\%$ & & $50.00\pm0.02$ & $1.000\pm 0.005$  & $50.00\pm 1.00$\\    
		        \hline
		        \ce{^{197}Au} & & $ < 0.100\%$ & & $30.00\pm0.02$  & $1.000\pm 0.005$  & $51.20\pm 0.70$ \\
	   		\hline \hline
			
	\end{tabular}
	\label{Tab.01}
\end{table*}

\section{Data Analysis }
\label{DataAnalysis}

The data analysis uses the pulse height weighting technique (PHWT). PHWT was originally applied by Macklin and Gibbons in the measurement of neutron capture cross section by the \ce{C_6F_6} detector \cite{bib.b6}. This method has shown suitable for the detector systems in Back-n beamline \cite{lxx.hjs,hxr.nst}. 

\begin{figure}[ht]
\centering
\includegraphics[width=.5\textwidth]{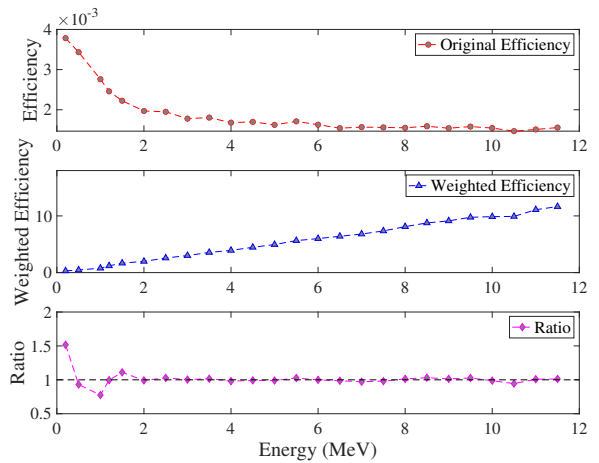}
\caption{(color online) (a) The \ce{C_6D_6} original efficiency. (b) Weighted efficiency. (c) The ratio of weighted efficiency to $\gamma$-ray energy. Below 1.5 MeV, the weighted efficiency is not ideally proportional to the energy, and hence a threshold is set here when processing the PH spectrum.}
\label{Fig.4}
\end{figure}

The experimental capture yields are obtained with a weighting function ($WF$) parameterized as a polynomial function of $\gamma$-ray energy. $WF$ can be expressed as
\begin{equation}\label{WF1}
WF(E_d) = \sum_{i=0}^{4}a_iE_d^i
\end{equation}
where $WF$ is the weight function, $E_d$ is an energy bin of pulse height spectrum, $a_i$ is the parameter of weight function and can be determined by the least squares fit.
\begin{equation}\label{WF1}
\chi^2 = \sum
\left (
E_{\gamma j}-\int_{EL}^{\infty}R(E_d,E_{\gamma j})WF(E_d)dE_d
\right )^2
\end{equation}
where $E_{\gamma j}$ is the $\gamma$-ray energy of group $j$, $R(E_d,E_{\gamma j})$ is the counts of pulse height (PH) spectrum with energy response function in $E_d$, $EL$ is the threshold of PH spectrum, $dE_d$ is the differentiation of $E_d$.
Each event is weighed by the proper $WF$ to ensure that the detector weighted efficiency is proportional to their excitation energy as shown in Fig. \ref{Fig.4}.

The cascaded $\gamma$-rays in ($n,\gamma$) reactions were simulated using the Geant4 \cite{bib.geant4} Monte-Carlo program. Since the simulation directly affects the uncertainty of the PHWT method, the target system and detector system are completely reconstructed as shown in Fig. \ref{Fig.021}(b) and (c). Here, the internally converted electron emission is also considered in the simulation. 

 The capture yield ($Yw$) can be determined using the following formula \cite{bib.yw}:
 \begin{equation}\label{Yw}
Y_w(E)=\frac{N_w}{N_sIS_n}
\end{equation}
where $N_w$ is the weighted PH spectrum count, $N_s$ is the sample area density, $S_n$ is the  neutron binding energy of the target, and $I$ is the neutron beam flux.

\begin{figure}[ht]
\centering
\includegraphics[width=.5\textwidth]{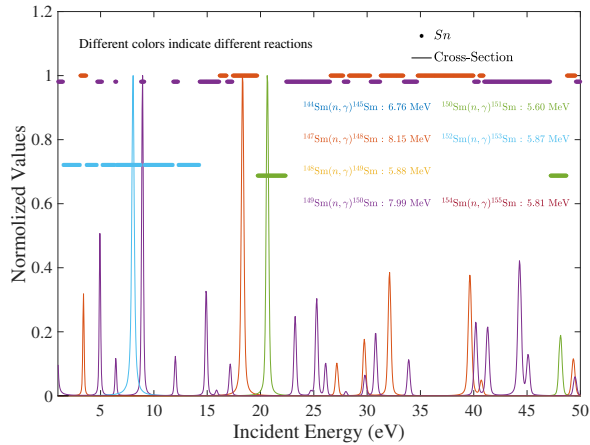}
\caption{(color online)  The values of $S_n$ at different resonance peaks, star point indicates normalized value of $S_n$ and solid line indicates normalized value of cross-section. Different colors indicate different isotopes. Note that both the $S_n$ and cross-section values are normalized to 0--1 for the convenience of showing the effect of the resonance peaks contributed by different isotopes on the value of $S_n$. }
\label{Fig.023}
\end{figure}

For an admixture of different isotopes, $S_n$ needs to be determined individually according to different resonance peaks, because each individual resonance belongs to one specific isotope and has its own separation energy for the capture efficiency. Fig. \ref{Fig.023} shows the values of $S_n$ at different resonance peaks. Star point indicates normalized value of $S_n$ and solid line indicates normalized value of cross-section. Different colors indicate different isotopes. Note that both the $S_n$ and cross-section values are normalized to 0--1 for the convenience of showing the effect of the resonance peaks contributed by different isotopes on the value of $S_n$.  Values of $S_n$ for different isotopes are shown in Table \ref{Tab.02}.

\begin{table}[ht]\small
	\caption{Values of $S_n$ for different isotopes }
	\begin{tabular}{c|c|c}
	   \hline \hline
	   Reaction & $S_n$ (MeV)  & Uncertainty ($\%$)\\ \hline
	   \ce{^{144}Sm}$(n,\gamma)$\ce{^{145}Sm} & $6.76$ & 0.003 \\ \hline
	    \ce{^{147}Sm}$(n,\gamma)$\ce{^{148}Sm} & $8.15$ & 0.002\\ \hline
	     \ce{^{148}Sm}$(n,\gamma)$\ce{^{149}Sm} & $5.88$ & 0.002\\ \hline
	      \ce{^{149}Sm}$(n,\gamma)$\ce{^{150}Sm} & $7.99$ & 0.002\\ \hline
	     \ce{^{150}Sm}$(n,\gamma)$\ce{^{151}Sm} & $5.60$ & 0.002\\ \hline
	     \ce{^{152}Sm}$(n,\gamma)$\ce{^{153}Sm} & $5.87$ & 0.002\\ \hline
	     \ce{^{154}Sm}$(n,\gamma)$\ce{^{155}Sm} & $5.81$ & 0.003\\ \hline
	     \hline
	   	\end{tabular}
	\label{Tab.02}
\end{table}

The uncertainty in the capture yield mainly includes the following factors \cite{lxx.prc}: uncertainty from experimental conditions and data analysis, and statistical error. 

Uncertainty from experimental conditions includes uncertainty of energy spectrum and the proton beam power. They directly affect the neutron flux in front of the target, and the uncertainty is passed into the yield through the $I$ term in Eq. (\ref{Yw}).  According to the Back-n collaboration \cite{cyh.epja}, the uncertainty of the energy spectrum in Back-n ES\#2 at the mode of without lead absorber is between 2.3\% and 4.5\%  above 0.15 MeV and less than 8.0\% below 0.15 MeV. Uncertainty from beam power is listed in Table \ref{Tab.01}.

Uncertainties in data analysis are mainly caused by the PHWT method and neutron binding energy of each reaction. The former will affect the uncertainty of the capture yield through $N_w$ in Eq.  (\ref{Yw}), while the latter will affect it through $S_n$ in Eq. (\ref{Yw}). 
In 2002, Tain et al. compared the neutron width PHWT treatment results of a 1.15 keV peak in $^{56}$Fe with experimental results, and found that the systematic error of PHWT was $2.00\%$--$3.00\%$ \cite{bib.phwt}. This uncertainty is only attainable if the proper threshold, conversion electron, and $\gamma$-ray summing effects are taken into account. We completely reconstructed the target system and detector system in the simulation. At the same time, a cascade $\gamma$ emission program including a model of internal conversion process is used in the simulation. These works can minimize the additional uncertainty when we applied PHWT to our results. The values of $S_n$ for different reactions are adopted from the new atomic mass evaluation  (AME2020) \cite{ame2020} and NuDat-3.0 database, each uncertainty is shown in Table \ref{Tab.02}. On the other hand, the uncertainty from the normalization method to determine the absolute value of $I$ in Eq. (\ref{Yw}) will also affect the capture yield uncertainty  The Ref. \cite{lxx.prc} provided two normalization methods:  Gaussian fitting of one of the resonance peak (usually the first peak in the experimental energy region is chosen, which for \ce{^{nat}Sm} target is the peak at 3.4 eV). The normalized coefficient is calculated by comparing the fitted curve with the evaluation data, and the CENDL-3.2 database was chosen in this work. Another method is to compare the energy bins one by one. For different targets, the value of normalized uncertainty is different, and for \ce{^{nat}Sm}, it is less than 1.3\%. 

\begin{figure}[ht]
\centering
\includegraphics[width=.5\textwidth]{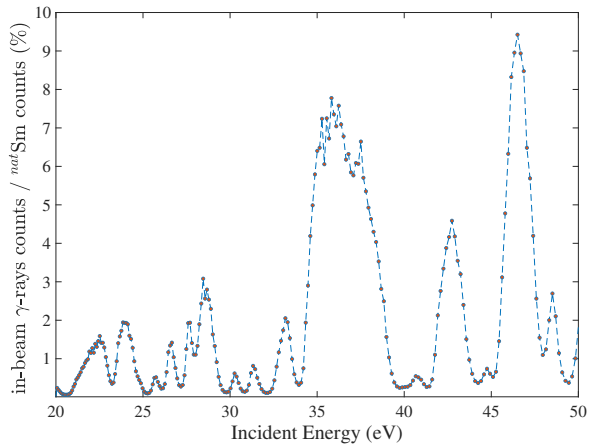}
\caption{(color online) Where the in-beam $\gamma$ background measured in Ref. \cite{lxx.cpb} was normalized to the neutron flux of this experiment. The ratio of \ce{^{nat}Sm} counts and in-beam $\gamma$-ray counts show the contribution of in-beam $\gamma$-ray background. The energy region is limited below 50 eV to control the uncertainty less than 10\%. According to the results of Ref. \cite{lxx.cpb}, the in-beam $\gamma$-ray background starts at 20 eV.}
\label{Fig.024}
\end{figure}

The in-beam $\gamma$-ray background at Back-n is a non-negligible noise component. Unfortunately, it was not fully understood in the early stages of the Back-n beamline experiment. After 2020, some in-beam background analyses at Back-n have been reported \cite{rj.nima,rj.aps}, and we also analyzed in-beam background in detail in the 2021 experiment \cite{lxx.cpb}.  Although the importance and measurement methods of the in-beam background are well understood, they are not effective in improving the experimental results in 2019 because of the changes in experimental conditions. Because there is no evidence that this does not have a significant effect on the in-beam $\gamma$ background. In addition, quantification of in-beam $\gamma$ background is also a problem. Based on the above reasons, the current feasible method is to select the part with little influence of the in-beam $\gamma$ background, and introduce it into the experimental results as uncertainty, as done in reference \cite{lxx.prc}. The cross-section of \ce{^{nat}Sm} is significantly lower than that of \ce{^{nat}Er}, which leads to a more significant effect of the in-beam $\gamma$ background. Therefore, we have to be careful to choose the lower energy region for analysis, because the in-beam $\gamma$-rays mainly exist in the energy region greater than 20 eV, and becomes larger with increasing neutron energy up to around 1 keV \cite{lxx.cpb}. The processing method is the same as in Ref. \cite{lxx.prc}, but the energy region is limited below 50 eV to control the uncertainty. Contribution of in-beam $\gamma$ background is shown in Fig. \ref{Fig.024} for the \ce{^{nat}Sm} target. 

Finally, the statistical uncertainty of the experiment was smaller than 0.68\%. All error sources and their estimates are summarized in Table \ref{Tab.03}.

\begin{table*}\small
	\caption{The statistical error and systematic error of the experiment}
	\begin{tabular}{c|c|c}
	   \hline\hline
	     $\sigma$ & Meaning & Value \\
	     \hline
	     \multicolumn{3}{c}{Experimental Conditions} \\
	     \hline
	     $\sigma({BeamPower})$ &Uncertainty from beam power & see Table \ref{Tab.01} \\
	     $\sigma({I_2})$ & Uncertainty from energy spectra below 0.15 MeV& $<8.00\%$ \\ 
	    \hline 
	     \multicolumn{3}{c}{Data Analysis} \\
	     \hline
	      $\sigma({PHWT})$ &Uncertainty from PHWT method& $<3.00\%$ \\
	      $\sigma({Normalized})$ &Uncertainty from normalized & $<1.30\%$ \\
	     $\sigma({In-Beam})$ & Uncertainty from counts of in-beam BKG& see Fig. \ref{Fig.023} \\
	      $\sigma({S_n})$ & Uncertainty from binding energy & see Table \ref{Tab.02} \\
	     \hline
	     \multicolumn{3}{c}{Statistical error} \\
	     \hline
	     $\sigma({Statistic})$ & Uncertainty from mathematical statistics& $<0.68\%$ \\ 
		\hline\hline
	\end{tabular}
	\label{Tab.03}
\end{table*}

\section{Result and Discussion}

We should admit that past analyses \cite{lxx.prc, lxx.cpb} are subject to unpredictable uncertainties. Because we used to deal with the cross-section directly according to the following Eq. \cite{bib.xc}:
\begin{equation}\label{Yw_XC}
Y_w(E) = (1-e^{Nt\sigma_t(E)})\frac{\sigma_c(E)}{\sigma_t(E)}
\end{equation}
where $\sigma_c$ is the neutron capture cross section, $\sigma_t$ is the neutron total cross section, $N$ is the atom density, and $t$ is the target thickness.  

In fact, this formula is ideal, which does not take into account the experimental effects. These neglected effects, such as Doppler broadening, energy resolution functions, etc., can cause errors in the experimental results. However, these uncertainties are difficult to assess.  Therefore, the cross-section is not directly calculated by Eq. \ref{Yw_XC} in this work. In order to obtain more accurate data we analyzed here, the capture yield by an $R-$Matrix \footnotesize{SAMMY} \normalsize code \cite{bib.sammy}.

In this work, we concerned about 
the neutron capture yield of natural samarium target measured in the resonance energy range of 1--50 eV. Fitting yields are more accurate than fitting cross-sectional data directly because various experimental effects, such as broadening, self-shielding, and multiple scattering, are explicitly considered by the \footnotesize{SAMMY} \normalsize  code.

Resolution broadening refers to the combined effects of the CSNS proton burst width, the time delay in the moderator, the TOF channel width, and the effect of the detector system. 
Resolution function of CSNS was studied in Ref. \cite{jb.nima}. Self-shielding correction accounts for the attenuation of the incident beam in the sample. Multiple-scattering correction accounts for the increase in the observed capture yield due to capture of neutrons scattered at higher energies.  Both corrections were applied in \footnotesize{SAMMY} \normalsize for all capture analyses. 

\begin{figure}[ht]
\centering
\includegraphics[width=.5\textwidth]{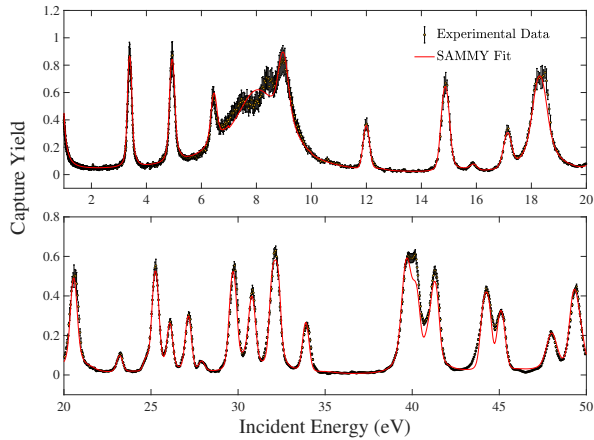}
\caption{(color online) The experimental capture yields fitted by the \footnotesize{SAMMY} \normalsize code. It is worth noting that for a 1-mm samarium target, the capture yield at 8 eV should saturate but in fact does not, and show signs of two peaks. Since it is currently uncertain whether these results arise from insufficient measurements or a new phenomenon, it is cautious to treat them as one peak. This requires further experimental verification. As the influence of the in-beam $\gamma$ background increases, the results for the energy region beyond 42 eV may have large errors, which brings challenges to the fitting. }
\label{Fig.025}
\end{figure}

The experimental yields and \footnotesize{SAMMY} \normalsize fitting results are shown in Fig. \ref{Fig.025}. It can be seen that for the resonance peaks in higher energy (above 42 eV), the fitting effect seems not very good, which may be caused by the gradual increase of the in-beam background. A probable contribution is the in-beam $\gamma$ background. According to the previous results \cite{lxx.cpb}, the contribution of the in-beam $\gamma$-rays increases between 20 eV and about 1 keV. It is worth noting that for a 1-mm samarium target, the capture yield at 8 eV should saturate (the value of the capture yield should reach 1) but in fact does not, and show signs of two peaks (see Fig. \ref{Fig.025}). At present, it cannot be directly concluded that there are two resonance peaks there, perhaps it is also due to the lack of measurement time. Therefore, they are treated cautiously as one peak. This requires further experimental verification. Resonance parameters are shown in Table 
\ref{Tab.04}.

\begin{table*}[htp]
	\centering
		\caption{Resonance parameters extracted from the $R$-matrix analysis and compared with they are from evaluation database of ENDF/B-VIII.0 and CENDL-3.2. As analyzed in text, the results of $E_R=8.0$ eV and $E_R>42$ eV were interfered by the shortage of measurement time and in-beam $\gamma$-rays, thus, these results have been removed.}
			\begin{tabular}{c|c|c|c|cc|cc|c|cc|cc}
				\hline\hline
				\multirow{2}*{Mass} &\multirow{2}*{J}  & \multirow{2}*{g} &\multirow{2}*{$E_R$(eV)}
				& \multicolumn{2}{c|}{ENDF/B-VIII.0} &  \multicolumn{2}{c|}{CENDL-3.2}  
				& \multicolumn{5}{c}{Present Work} \\ \cline{5-13}
				&  & &
				& $\Gamma_\gamma$ (meV) & $\Gamma_n$ (meV) 
				&  $\Gamma_\gamma$ (meV) & $\Gamma_n$ (meV) 
				& $E_R$(eV)
				& $\Gamma_\gamma$ (meV)  & $std(\Gamma_\gamma)$ (meV)  
				& $\Gamma_n$ (meV) & $std(\Gamma_n)$ (meV) \\ \hline
\multirow{7}*{147}
&\multirow{4}*{3.0}	&\multirow{4}*{0.44}	
&	3.40	&	65.00	&	1.30	&	67.00	&	1.40	&	3.39	&	48.48	&	2.81	&	1.17	&	0.04	\\ \cline{4-13}
&	&	&	27.16	&	79.00	&	6.80	&	69.00	&	7.00	&	27.14	&	58.12	&	3.69	&	5.67	&	0.12	\\ \cline{4-13}
&	&	&	29.74	&	73.00	&	14.70	&	71.00	&	14.70	&	29.73	&	93.32	&	3.37	&	19.64	&	0.49	\\ \cline{4-13}
&	&	&	40.71	&	73.40	&	5.10	&	69.00	&	5.40	&	40.80	&	45.18	&	4.07	&	4.26	&	0.31	\\ \cline{2-13}
&\multirow{3}*{4.0}	&\multirow{3}*{0.56}
&	18.32	&	70.00	&	69.30	&	72.00	&	71.90	&	18.31	&	309.56	&	11.84	&	19.05	&	0.48	\\ \cline{4-13}
&	&	&	32.10	&	68.00	&	38.20	&	62.00	&	39.00	&	32.11	&	41.70	&	1.27	&	55.05	&	1.42	\\ \cline{4-13}
&	&	&	39.64	&	67.00	&	70.20	&	60.00	&	71.30	&	39.72	&	292.67	&	4.38	&	30.08	&	0.57	\\ \cline{1-13}
\multirow{17}*{149}
&\multirow{7}*{3.0}	&\multirow{7}*{0.44}	
&	6.43	&	66.00	&	0.80	&	68.00	&	1.20	&	6.46	&	116.01	&	10.09	&	1.55	&	0.09	\\ \cline{4-13}
&	&	&	12.01	&	65.50	&	1.90	&	62.00	&	1.80	&	11.99	&	5.01	&	0.46	&	2.88	&	0.25	\\ \cline{4-13}
&	&	&	15.88	&	85.00	&	0.40	&	62.00	&	0.40	&	15.89	&	48.96	&	5.08	&	0.35	&	0.02	\\ \cline{4-13}
&	&	&	25.27	&	61.50	&	15.70	&	69.00	&	16.00	&	25.25	&	89.04	&	4.99	&	16.46	&	0.44	\\ \cline{4-13}
&	&	&	28.01	&	40.00	&	0.50	&	62.00	&	0.60	&	27.92	&	26.69	&	2.75	&	0.78	&	0.04	\\ \cline{4-13}
&	&	&	29.80	&	62.00	&	3.30	&	62.00	&	3.20	&	30.30	&	94.66	&	9.33	&	0.80	&	0.06	\\ \cline{4-13}
&	&	&	40.20	&	56.00	&	26.70	&	62.00	&	28.30	&	40.21	&	66.00	&	5.86	&	20.92	&	0.88	\\ \cline{2-13}
&\multirow{10}*{4.0}	&\multirow{10}*{0.56}
&	0.87	&	60.80	&	0.70	&	62.70	&	0.80	&	0.84	&	92.55	&	6.81	&	0.73	&	0.05	\\ \cline{4-13}
&	&	&	4.94	&	64.00	&	1.90	&	59.00	&	2.10	&	4.94	&	58.58	&	4.14	&	1.65	&	0.07	\\ \cline{4-13}
&	&	&	8.93	&	66.50	&	9.50	&	70.00	&	11.80	&	8.98	&	196.42	&	15.55	&	8.42	&	0.55	\\ \cline{4-13}
&	&	&	14.91	&	66.00	&	5.60	&	65.00	&	5.60	&	14.87	&	44.70	&	2.78	&	6.74	&	0.27	\\ \cline{4-13}
&	&	&	17.16	&	89.00	&	2.00	&	62.00	&	2.00	&	17.13	&	84.48	&	5.98	&	2.43	&	0.08	\\ \cline{4-13}
&	&	&	23.25	&	72.00	&	7.90	&	62.00	&	0.90	&	23.20	&	151.06	&	7.40	&	1.21	&	0.02	\\ \cline{4-13}
&	&	&	24.72	&	40.00	&	0.30	&	62.00	&	0.30	&	24.71	&	43.10	&	4.16	&	0.63	&	0.03	\\ \cline{4-13}
&	&	&	26.11	&	49.00	&	3.30	&	62.00	&	3.20	&	26.08	&	35.09	&	3.32	&	3.46	&	0.08	\\ \cline{4-13}
&	&	&	30.80	&	66.00	&	9.40	&	60.00	&	9.50	&	30.81	&	85.72	&	5.47	&	8.77	&	0.16	\\ \cline{4-13}
&	&	&	33.90	&	55.00	&	5.60	&	62.00	&	5.60	&	33.88	&	5.27	&	0.40	&	5.10	&	0.12	\\ \cline{1-13}
150
&	0.5 &	1.00
&	20.65	&	60.20	&	47.90	&	60.20	&	47.70	&	20.53	&	176.34	&	5.22	&	9.42	&	0.17	\\ 
\hline\hline
	\end{tabular}
	\label{Tab.04}
\end{table*}

The resonance parameters are used to calculate the single-level and multi-level Breit-Wigner cross-sections. For both single-level and multi-level cases, the capture cross section is given by \cite{SLBW}
 \begin{equation}\label{sigma}
\sigma_{\ce{capture}} = \frac{\pi}{k^2} \sum_J g_J \sum_c \sum_\lambda \frac{\Gamma_{\lambda c} \Gamma_{\lambda \gamma}}{D_\lambda}
\end{equation}
where $k$ is the center-of-mass momentum, $\Gamma_{c}$ is the resonance width of incident channels, $\Gamma_{\gamma}$ is the $\gamma$ width, $g_J$ is the statistical $g$ factor given by
 \begin{equation}\label{g}
g_J = \frac{2\boldsymbol{J}+1}{(2\boldsymbol{s}+1)(2\boldsymbol{I}+1)}
\end{equation}
where $\boldsymbol{s} = 1/2$ is the neutron spin, $\boldsymbol{I}$ is the ground state spin of the target nucleus. Here $\boldsymbol{J}=|\boldsymbol{L}+\boldsymbol{S} |$, $\boldsymbol{S}=\boldsymbol{s}+\boldsymbol{I}$, with $\boldsymbol{L}$ the orbital angular momentum. The denominator $D_\lambda$ in Eq. (\ref{sigma}) represents
 \begin{equation}\label{D}
D_\lambda = (E-E_{\lambda})^2 + (\Gamma_\gamma/2)^2
\end{equation}
where $E$ is the center-of-mass incident energy, $E_{\lambda}$ is the resonance energy, and $\Gamma_\gamma = \sum_c \Gamma_{\lambda c} + \Gamma_{\lambda \gamma}$. 

\begin{figure*}[ht]
\centering
\includegraphics[width=1.\textwidth]{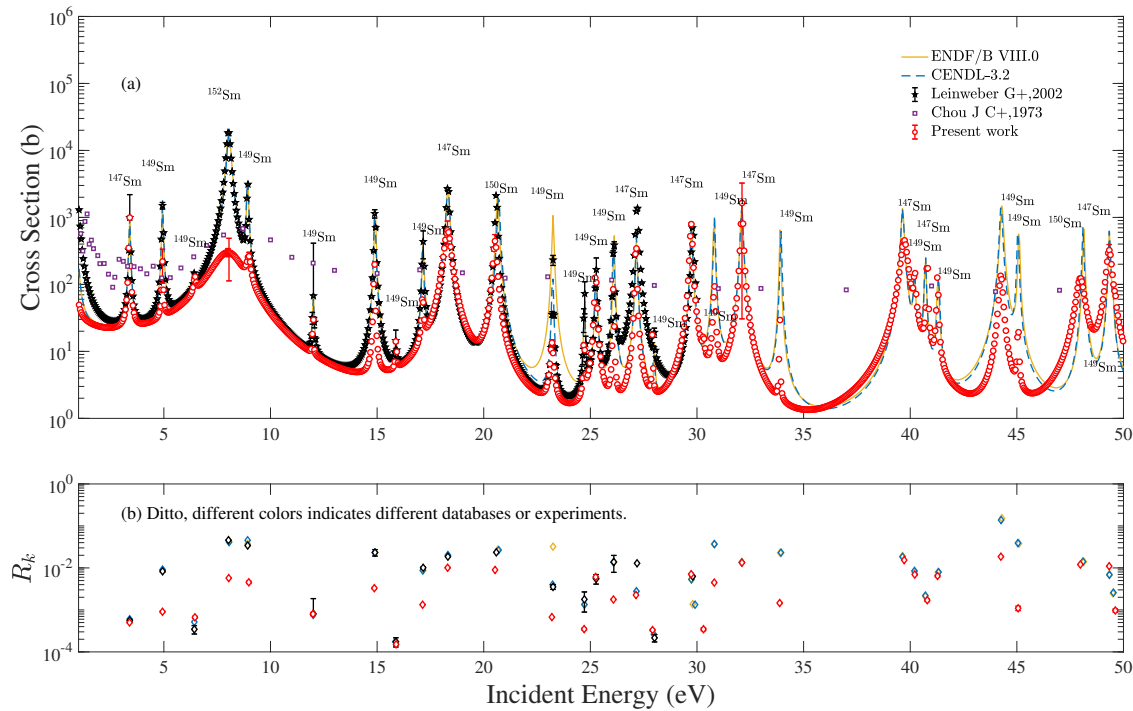}
\caption{(color online)  (a) Comparison of theoretical cross-section calculated by resonance parameters with other experimental data and evaluation databases of ENDF/B-VIII.0 and CENDL-3.2. It should be noted that reference \cite{bib.gl} does not directly provided the absolute cross section, this work calculates the theoretical cross section by applying their resonance parameters  to Eq. (\ref{sigma}). A clear difference between ENDF/B VIII.0 and CENDL-3.2 database appears at 23.2 eV, the existing experimental results may be more in consistent with CENDL-3.2 database. Most of the controversial experimental results invariably come from the samarium 149 isotope. (b) $R_k$ values of different experiments and evaluation databases, different colors represent different experimental and evaluation results. }
\label{Fig.026}
\end{figure*}

The resonance parameters obtained in this work are applied to Eq. \ref{sigma}, and the present cross sections are calculated as shown in Fig. \ref{Fig.026}(a). The existing experimental data and the evaluation database of ENDF/B-VIII.0 and CENDL-3.2 are also shown for conparision. It should be noted that Ref. \cite{bib.gl} did not directly provided the absolute cross section, we calculated the cross sections by using their resonance parameters. To better evaluate the effect of SAMMY fitting, resonance kernels $R_k$ for different experiments and evaluation databases are calculated by the following Eq. (\ref{Rk}) \cite{wdx.aps} and displayed in Fig. \ref{Fig.026}(b).
 \begin{equation}\label{Rk}
R_k = g_J \Gamma_\gamma \Gamma_n/(\Gamma_n+\Gamma_\gamma)
\end{equation}

Fig. \ref{Fig.025} can explain the strange result at 8 eV in Fig. \ref{Fig.026}. From the results of the capture yield, the resonance peak of the samarium 152 isotope at 8 eV may be contributed by two small resonance peaks. Another possible reason is the lack of measurement time.  For 1-mm thick samarium target, the neutron capture yield should saturate here, but it does not.  Since the measurement of the samarium target is scheduled at the end, the measurements of the background and the gold target squeezed the time that should belong to samarium target.  A hindsight experience is that gold target doesn't need to be paid so much attention as a test. Therefore, we cannot completely rule out a cause from measurements, and hence we recommend a further experiment to check this result.

The experimental results are good between 20 and 42 eV according to Fig. \ref{Fig.026}. At 23.2 eV, there is a clear divergence between the ENDF/B VIII.0 and CENDL-3.2 databases, and it also can be seen form the results of $R_k$ in Fig. \ref{Fig.026}(b). The results of both experiments suggest that the resonance peak at 23.2 eV contributed by samarium 149 isotope may be lower than ENDF/B-VIII.0 database,  but close to CENDL-3.2 database. At 27.1 eV, our $R_k$ values are significantly smaller than the results of Leinweber et al. but evaluation database.
 
At 30-50 eV, only one set of published experimental data exists and cannot be used to evaluate resonances \cite{bib.cjc}. The resonance at 34 eV is different, and the value of $R_k$ provided by the experiment is significantly smaller than that of the evaluation databases. It may not come from background or statistics. The $R_k$ results for the peaks above 42 eV are significantly smaller than evaluation database. Because of a significant increase in the in-beam background, we can't rule out the reasons for the experiment. Thus, we plan to conduct more experiments to determine it.

Interestingly, most of the controversial experimental results invariably come from the samarium 149 isotope. It is the fission product of \ce{^{235}U}, and its neutron capture cross section plays an important role in the design and construction of nuclear reactors.  According to the EXFOR database, few experiments have focused on the neutron capture cross-section of \ce{^{149}Sm} target in the resonance energy region. It deserves further study.

\section{summary and conclusions}
\label{summaryhan}

We have measured the neutron capture cross yield of the natural samarium target in the 1--50 eV energy region at the Back-n beamline of CSNS. Resonance parameters are extracted and cross sections are calculated accordingly. We found a clear difference between ENDF/B VIII.0 and CENDL-3.2 database at 23.2 eV, the existing experimental results may be more in consistent with CENDL-3.2 database. The difference between different experiments is obvious at 27.1 eV, however, the two evaluation databases are almost identical here. The experimental results at 30-50 eV are new data and mostly consistent with the evaluation database except for the resonance peak at 34 eV. The resonance width analyzed from the experimental data is significantly smaller than that from evaluation database. The reason is less likely due to background or statistics. 

Two important lessons learned from this experiment. One is about the time allocation of measurements for gold target and background, especially for gold target. The gold target as the main target was fully measured in January 2019, and the result was published elsewhere \cite{lxx.hjs, hxr.nst}. In subsequent experiments, the gold targets were only measured as validation data to ensure the correctness of each experimental setup and data analysis. Exact data are not necessary, so it shouldn't take too much beam time. On the other hand, the in-beam $\gamma$ background has always been a priority issue. It directly determines the accuracy of the data and limits the energy range that can be reported. We have published data from 1-100 eV in the analysis of Er targets \cite{lxx.prc}, and data only from  1 to 50 eV can be considered reliable and are provided in this work. Because for higher energy region, the influence of the in-beam $\gamma$ background may greater than $10\%$. Although we did detailed measurement and analysis later \cite{lxx.cpb}, this is not universal. Fortunately, related work is underway. More precise data is to be expected.

\begin{acknowledgments}

We appreciate effective technical support from Dr. Yi-Jie Wang at Tsinghua University, Dr. Yu-Chao Xu at General Electric, and the efforts of the staff of the CSNS and Back-n collaboration.
This work was supported by the National Natural Science Foundation of China under Grant Nos. 11875311, 11905274, 1705156, U2032146, 11865010, 11765015, and 1160509, Natural Science Foundation of Inner Mongolia under Grant Nos. 2019JQ01 and 2018MS01009, and the Strategic Priority Research Program of the CAS (No. XDB34030000).

\end{acknowledgments}


\end{CJK*} 

\end{document}